\documentclass[10pt,a4paper,final,twocolumn]{article}

\pdfoutput=1

\usepackage{amssymb}
\usepackage[english]{babel}
\usepackage{color}
\usepackage{enumerate}
\usepackage[breaklinks=true,%
            colorlinks=false,%
            pdfkeywords={OCaml, Just In Time compilation, machine code generation},%
            pdftitle={Towards a native toplevel for the OCaml language},%
            pdfauthor={Marcell Fischbach, Benedikt Meurer},%
            pdfsubject={},%
            pdfdisplaydoctitle=true]{hyperref}
\usepackage{tikz}
\usepackage{varwidth}
\usepackage{xspace}

\newcommand{\ie}{i.e.\@\xspace}

\usetikzlibrary{arrows}
\usetikzlibrary{trees}
\usetikzlibrary{arrows,backgrounds,decorations.pathmorphing,fit,matrix,positioning}
\tikzset{
  phase/.style={
      rectangle,
      inner sep=1mm,
      rounded corners=1mm,
      minimum size=6mm,
      very thick,
      draw=blue!50!black!50,
      top color=white,
      bottom color=blue!50!black!20
  }
}

\definecolor{ocamljit2}{HTML}{FF0000}
\definecolor{ocamlnatext}{HTML}{00FF00}
\definecolor{ocamlnatjit}{HTML}{0000FF}

\begin{document}

\title{%
  Towards a native toplevel for the OCaml language
}
\author{%
  Marcell Fischbach \\
  Compilerbau und Softwareanalyse \\
  Fakult\"at IV, Universit\"at Siegen \\
  D-57068 Siegen, Germany \\
  \url{marcellfischbach@googlemail.com}
  \and
  Benedikt Meurer\thanks{Corresponding Author} \\
  Compilerbau und Softwareanalyse \\
  Fakult\"at IV, Universit\"at Siegen \\
  D-57068 Siegen, Germany \\
  \url{meurer@informatik.uni-siegen.de}
}
\date{}

\maketitle

\begin{abstract}
  This paper presents the current state of our work on an interactive toplevel for the OCaml language
  based on the optimizing native code compiler and runtime. Our native toplevel is up to $100$ times
  faster than the default OCaml toplevel, which is based on the byte code compiler and interpreter.
  It uses Just-In-Time techniques to compile toplevel phrases to native code at runtime, and
  currently works with various Unix-like systems running on x86 or x86-64 processors.
\end{abstract}

\section{Introduction}

The OCaml \cite{Leroy11,Remy02} system is the main implementation of the Caml
language \cite{Caml11}, featuring a powerful module system
combined with a full-fledged object-oriented layer. It ships with an optimizing native
code compiler \texttt{ocamlopt}, for high performance; a byte code compiler \texttt{ocamlc}
and interpreter \texttt{ocamlrun}, for increased portability; and an interactive top-level
\texttt{ocaml} based on the byte code compiler and runtime, for interactive use of OCaml
through a read-eval-print loop.

\texttt{ocamlc} and \texttt{ocaml} translate the source code into a sequence of byte code
instructions for the OCaml virtual machine \texttt{ocamlrun}, which is based on the ZINC
machine \cite{Leroy90} originally developed for Caml Light \cite{Leroy02}. The optimizing
native code compiler \texttt{ocamlopt} produces fast machine code for the supported targets
(at the time of this writing, these are Alpha, ARM, Itanum, Motorola 68k, MIPS, PA-RISC, PowerPC,
Sparc, and x86/x86-64), but is currently only applicable to \emph{static program compilation}.
For example, it cannot yet be used with multi-stage programming in MetaOCaml \cite{Taha03,Taha06},
or the interactive toplevel \texttt{ocaml}.

This paper presents our work\footnote{The initial work in this area was done as part of the first
author's diploma thesis.} on a new native OCaml toplevel, called \texttt{ocamlnat}, which is
based on the native runtime, the compilation engine of the optimizing native code compiler and
an earlier prototype implementation of a native toplevel by Alain Frisch. Our implementation
currently supports x86 and x86-64 processors \cite{Amd09Vol1,Intel10Vol1} and should work with any
POSIX compliant operating system supported by the OCaml native code compiler. It is verified to
work with Mac OS X 10.6 and 10.7, Debian GNU/Linux 6.0 and above, and CentOS 5.6 and 5.7. The
full source code is available from the \texttt{ocamljit-nat} branch of the \texttt{ocaml-experimental}
Git repository hosted on GitHub at \cite{Meurer11ocamlexperimental}.

The paper is organized as follows: Section~\ref{section:Motivation} motivates the need for a
usable native OCaml toplevel. Section~\ref{section:Overview_of_the_OCaml_compilers} presents
an overview of the OCaml compilers and Section~\ref{section:The_native_toplevel} describes the
previous \texttt{ocamlnat} prototype which inspired our work, while
Section~\ref{section:Just_In_Time_code_generation} presents our work on \texttt{ocamlnat}.
Performance measures are given in Section~\ref{section:Performance}.
Sections~\ref{section:Related_and_further_work} and \ref{section:Conclusion} conclude with
possible directions for further work.

\section{Motivation} \label{section:Motivation}

Interactive toplevels are quite popular among dynamic and scripting languages like Perl, Python, Ruby
and Shell, but also with functional programming languages like OCaml, Haskell and LISP. In case of
scripting languages the interactive toplevel is usually the only frontend to the underlying interpreter
or Just-In-Time compiler.

In case of OCaml, the interactive toplevel is only one possible interface to the byte code interpreter;
it is also possible to separately compile source files to byte or native code object files, link them
into libraries or executables,
and deploy these libraries or executables. The OCaml toplevel is therefore mostly used for
interactive development, rapid prototyping, teaching and learning, as an interactive program console
or for scripting purposes.

The byte code runtime is the obvious candidate to drive the interactive toplevel, because the platform
independent byte code is very portable and easy to generate -- compared to native machine code. And in
fact the byte code toplevel has served users and developers well during the last years. But nevertheless
there are valid reasons to have a native code toplevel instead of or in addition to the byte code toplevel:

\paragraph{Performance}

This is probably the main reason why one wants to have a native code toplevel. While the performance of
the byte code interpreter is acceptable in many cases (which can be improved by using one of the
available Just-In-Time compilers \cite{Meurer10ocamljit,Meurer10jit,Meurer11ocamljit2,Starynkevitch04}),
it is not always sufficient to handle the necessary computations. Sometimes one needs the execution
speed of the native runtime, which can be up to hundred times faster than the byte code runtime as we will
show in Section~\ref{section:Performance}.

For example, the Mancoosi project \cite{Mancoosi11} has developed a library that allows to perform analysis
of large sets of packages in free software distributions, that can be done acceptably efficiently with the
native code compiler and runtime, but are too slow in bytecode. To perform interactive analysis (\ie select
packages with particular properties, analyse them, \ldots), having a native toplevel is really the only way
to go for them, as it can combine the flexibility of the toplevel interaction with the speed of native code.

Tools such as ocamlscript \cite{ocamlscript11} try to combine the performance of the code generated by
the optimizing native code compiler with the flexibility of a ``scripting language interface''. But this
is basically just a work-around -- with several limitations. A native toplevel would address this issue in
a much cleaner and simpler way.

\paragraph{Native runtime}

There are scenarios where only the native code runtime is available and hence the byte code toplevel,
which depends on the byte code runtime, cannot be used. One recent example here is the Mirage cloud
operating system \cite{Mirage11,Madhavapeddy10,Madhavapeddy10hotcloud}, which compiles OCaml programs
to Xen micro-kernels \cite{Barham03} and executes them via the Xen hypervisor \cite{Xen11}. Mirage
uses the OCaml toplevel as OS console, but is currently limited to the byte code toplevel in read-only
mode, due to the lack of a toplevel that works with the native code runtime.

\section{Overview of the OCaml compilers} \label{section:Overview_of_the_OCaml_compilers}

In this section we briefly describe the OCaml compilers, covering both the byte code compiler
\texttt{ocamlc} and the optimizing native code compiler \texttt{ocamlopt}.
Feel free to skip to section~\ref{section:The_native_toplevel} if you are already familiar with the details.

\begin{figure}[htb]
  \centering
  \begin{tikzpicture}[node distance=8mm,text height=1.5ex,text depth=.25ex]
    \matrix[row sep={1.3cm,between origins}]
    {
      & \node (start) {}; \\
      & \node[phase] (parsing) {Parsing}; \\
      & \node[phase] (typing) {Typing}; \\
      & \node[phase] (transl) {Translate}; \\
      & \node[phase] (simplif) {Simplify}; \\
      \node[phase] (bytegen) {Bytegen}; && \node[phase] (asmgen) {Asmgen}; \\
      \node (bytecode) {}; && \node (nativecode) {}; \\
    };
    \draw[->] (start) -- node[right] {\it source program} (parsing);
    \draw[->] (parsing) -- node[right] {\tt Parsetree} (typing);
    \draw[->] (typing) -- node[right] {\tt Typedtree} (transl);
    \draw[->] (transl) -- node[right] {\tt Lambda} (simplif);
    \draw[->] (simplif) -- node[left] {} (bytegen);
    \draw[->] (simplif) -- node[right] {} (asmgen);
    \draw[->] (bytegen) -- node[left] {\it byte code} (bytecode);
    \draw[->] (asmgen) -- node[right] {\it native code} (nativecode);
  \end{tikzpicture}
  \caption{The OCaml compilers}
  \label{fig:The_OCaml_compilers}
\end{figure}
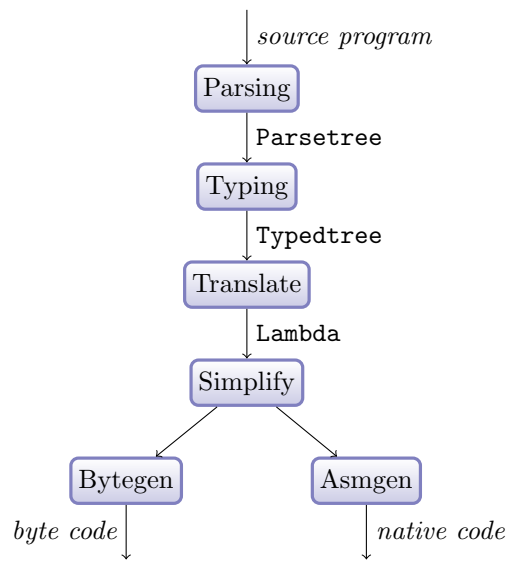

Figure~\ref{fig:The_OCaml_compilers} gives an overview of the compiler phases
and representations in the OCaml byte and native code compilers. Compilation always starts
by parsing an OCaml source program (either from a source file or a source region in
interactive mode) into an abstract syntax tree (AST, see file \texttt{parsing/parsetree.mli}
of the OCaml source code). Compilation then proceeds by computing the type annotations to
produce a typed syntax tree (see file \texttt{typing/typedtree.mli}).

From this typed syntax tree, the compiler generates a so called \emph{lambda representation} (see file
\texttt{bytecomp/lambda.mli}) inspired by the untyped call-by-value $\lambda$-calculus
\cite{Appel98ml,Jones87,Michaelson89}. This lambda representation is then optimized by
transforming lambda trees into \emph{better} or smaller lambda trees (see file
\texttt{bytecomp/simplif.ml}), yielding a final platform independent, internal
representation of the source program as result of the compiler frontend phases.

The simplified lambda representation is then used as input for the respective compiler
backend, which is either
\begin{itemize}
\item the \texttt{Bytegen} module in case of the byte code compiler
  (see file \texttt{bytecomp/bytegen.ml}), or
\item the \texttt{Asmgen} module in case of the optimizing native code compiler
  (see file \texttt{asmcomp/asmgen.ml}).
\end{itemize}

The byte code backend, which is used by the byte code compiler \texttt{ocamlc} as well as
the byte code toplevel \texttt{ocaml}, basically transforms the simplified lambda representation
into an equivalent byte code program (see file \texttt{bytecomp/instruct.mli}), suitable for
(a) direct execution by the byte code interpreter \texttt{ocamlrun} or (b) just-in-time
compilation using either OCAMLJIT \cite{Starynkevitch04} or OCamlJIT2
\cite{Meurer10jit,Meurer10ocamljit,Meurer11ocamljit2}.
This is done by the \texttt{Emitcode} module (see file \texttt{bytecomp/emitcode.ml}).
Additional details about the byte code compiler and runtime can be found in \cite{Leroy90},
\cite{Meurer10ocamljit} and \cite{Starynkevitch04}.

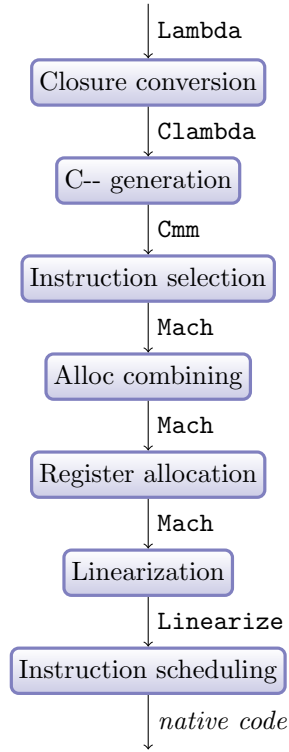
\begin{figure}[htb]
  \centering
  \begin{tikzpicture}[node distance=8mm,text height=1.5ex,text depth=.25ex]
    \matrix[row sep={1.3cm,between origins}]
    {
      \node (start) {}; \\
      \node[phase] (closure) {Closure conversion}; \\
      \node[phase] (cmmgen) {C\mbox{-}\mbox{-} generation}; \\
      \node[phase] (selection) {Instruction selection}; \\
      \node[phase] (comballoc) {Alloc combining}; \\
      \node[phase] (regalloc) {Register allocation}; \\
      \node[phase] (linearize) {Linearization}; \\
      \node[phase] (scheduling) {Instruction scheduling}; \\
      \node (end) {}; \\
    };
    \draw[->] (start) -- node[right] {\tt Lambda} (closure);
    \draw[->] (closure) -- node[right] {\tt Clambda} (cmmgen);
    \draw[->] (cmmgen) -- node[right] {\tt Cmm} (selection);
    \draw[->] (selection) -- node[right] {\tt Mach} (comballoc);
    \draw[->] (comballoc) -- node[right] {\tt Mach} (regalloc);
    \draw[->] (regalloc) -- node[right] {\tt Mach} (linearize);
    \draw[->] (linearize) -- node[right] {\tt Linearize} (scheduling);
    \draw[->] (scheduling) -- node[right] {\it native code} (end);
  \end{tikzpicture}
  \caption{Native code generation (\texttt{Asmgen} module)}
  \label{fig:Native_code_generation}
\end{figure}

The native code backend, which is used by the optimizing native code compiler \texttt{ocamlopt}
as well as the native toplevel \texttt{ocamlnat}, is shown in Figure~\ref{fig:Native_code_generation}.
It takes the simplified lambda representation as input and starts by transforming it into a variant
of the lambda representation (see file \texttt{asmcomp/clambda.mli}) with explicit closures and
explicit direct/indirect function calls (see file \texttt{asmcomp/closure.ml}). This is then further
processed and transformed into an equivalent representation in an internal dialect of C\mbox{-}\mbox{-}
\cite{JonesR98,JonesRR99} (see files \texttt{asmcomp/cmm.mli} and \texttt{asmcomp/cmmgen.ml}).

Afterwards the Instruction selection phase (see file \texttt{asmcomp/selection.ml}) picks appropriate
instructions for the target machine, transforming the C\mbox{-}\mbox{-} code into a tree based representation
of the machine code (see file \texttt{asmcomp/mach.mli}). The next step attempts to combine multiple heap
allocations within a basic block into a single heap allocation (see file \texttt{asmcomp/comballoc.ml}),
prior to allocating and assigning physical registers to the virtual registers used in the machine code
(see function \texttt{regalloc} in file \texttt{asmcomp/asmgen.ml}).
The final phases linearize the machine code (see file \texttt{asmcomp/linearize.ml}) and perform
instruction scheduling for better performance (see file \texttt{asmcomp/scheduling.ml}), yielding the
final representation of the (linearized) machine code.

The optimizing native code compiler \texttt{ocamlopt} writes the linearized machine code output of the
\texttt{Asmgen} module to an assembly file in the appropriate format for the target platform (see file
\texttt{asmcomp/emit.ml}), \ie using AT\&T assembly syntax on Linux and Mac OS X while using Intel assembly
syntax on Windows, and invokes the assembler from the system compiler toolchain, \ie GNU \texttt{as} on Linux,
to generate an object file. This object file can then be linked with other OCaml modules and C code into
an executable binary or a dynamic library file.

\section{The native toplevel} \label{section:The_native_toplevel}

In 2007 Alain Frisch added support for the \texttt{Dynlink} library to the native code compiler and
runtime, which was first made available as part of OCaml 3.11. This change made it possible to use
the OCaml native code runtime with dynamically loaded plugins, a feature that was previously only
available with the byte code runtime. Besides various other benefits, this also made it possible
to reuse the existing functionality of the optimizing native code compiler within the scope of a
native toplevel.

The initial proof-of-concept prototype of a native toplevel, developed by Alain Frisch and named
\texttt{ocamlnat}, was since then silently shipped with every OCaml source code release\footnote{It
must be build explicitly using \texttt{make ocamlnat} after \texttt{make world} and \texttt{make opt},
and it is only available for targets that support the native \texttt{Dynlink} library.}.

\begin{figure}[htb]
  \centering
  \begin{tikzpicture}[node distance=8mm,text height=1.5ex,text depth=.25ex]
    \matrix[row sep={1.3cm,between origins}]
    {
      \node (start) {}; \\
      \node[phase] (compiler) {Native code compiler}; \\
      \node[phase] (emit) {Native code emitter}; \\
      \node[phase] (as) {Toolchain Assembler (\texttt{as})}; \\
      \node[phase] (ld) {Toolchain Linker (\texttt{ld})}; \\
      \node[phase] (rtld) {Runtime Linker}; \\
      \node (end) {}; \\
    };
    \draw[->] (start) -- node[right] {\it OCaml phrase} (compiler);
    \draw[->] (compiler) -- node[right] {\it native code} (emit);
    \draw[->] (emit) -- node[right] {\it assembly file} (as);
    \draw[->] (as) -- node[right] {\it object file} (ld);
    \draw[->] (ld) -- node[right] {\it dynamic library file} (rtld);
    \draw[->] (rtld) -- node[right] {\it executable code} (end);
  \end{tikzpicture}
  \caption{\texttt{ocamlnat} prototype}
  \label{fig:ocamlnat_prototype}
\end{figure}

Figure~\ref{fig:ocamlnat_prototype} gives an overview of the internals of this \texttt{ocamlnat} prototype.
It works by starting up the OCaml native runtime and then prompts the user for OCaml phrases to evaluate
(just like the byte code toplevel \texttt{ocaml} does). Whenever the user enters a phrase, it is compiled
to native code using the modules of the optimizing native code compiler (utilizing the frontend phases as
shown in Figure~\ref{fig:The_OCaml_compilers} and the native backend phases as shown in
Figure~\ref{fig:Native_code_generation}).

This \emph{native code} is written to a temporary \emph{assembly file} by the Native code emitter, which
is also part of \texttt{ocamlopt}. The \emph{assembly file} is then passed to the Toolchain Assembler, \ie GNU
\texttt{as} on Linux, to produce a temporary \emph{object file}. This \emph{object file} is afterwards turned
into a \emph{dynamic library file} by the Toolchain Linker, \ie GNU \texttt{ld} on Linux, and loaded into the
native toplevel process using the Runtime Linker, finally yielding a memory area with the \emph{executable code}
which is then executed.

While this approach has the immediate benefit of requiring only a few hundred lines of glue code to turn
the existing modules of the optimizing native code compiler and the native \texttt{Dynlink} library into
a native toplevel, there are also several obvious drawbacks to this approach -- preventing wide-spread
adoption of \texttt{ocamlnat}:

\paragraph{Dependency on the system toolchain}

This is the most important problem of the native toplevel prototype as it prevents from being
used in areas that would really benefit from a native toplevel but do not have the toolchain programs
available. For example, the Mirage cloud operating system \cite{Mirage11,Madhavapeddy10,Madhavapeddy10hotcloud}
compiles OCaml programs to Xen micro-kernels, which are then executed by the Xen hypervisor \cite{Xen11}; Mirage uses
the OCaml toplevel as OS console, but is limited to the byte code toplevel in read-only mode right now, as
there is obviously no GNU toolchain available in a Xen micro-kernel.

It is worth noting that the toolchain dependency is also a problem with the optimizing native code compiler
\texttt{ocamlopt} on certain platforms such as Microsoft Windows where it is often a non-trivial task to
install the system toolchain. This is one of the reasons why companies such as LexiFi provide custom OCaml
distributions with an integrated toolchain.

\paragraph{Latency}

While the latency caused by reading and writing the assembly, object and library files as well as invoking the
external toolchain programs is not necessarily a show-stopper for an interactive toplevel, it is nevertheless
quite noticable, especially with short running programs or programs with many phrases, as we will see in
Section~\ref{section:Performance}.

\paragraph{Temporary library files}

On Microsoft Windows it is impossible to delete a library file that is currently loaded into a process, which
means that the prototype ``leaks'' one library file per toplevel phrase.

\paragraph{Unclear maintenance status}

Many people don't even know about \texttt{ocamlnat}, and those who do cannot rely on it. This is because
\texttt{ocamlnat} is not part of an OCaml installation, even though it ships as part of the source code
distribution, and it is not documented anywhere.

This is not so much a technical argument against the current approach, but it highlights its status as
being a proof-of-concept with no clear direction from the users point of view.

\section{Just-In-Time code generation} \label{section:Just_In_Time_code_generation}

We aim to improve \texttt{ocamlnat} in a way that avoids the drawbacks of the earlier prototype
and turns the native toplevel into a viable alternative to the byte code toplevel. As noted above,
the major drawback of Alain's prototype is the dependency on the system toolchain, that is, the
external assembler and linker programs.

Therefore we had to replace the last four phases of the \texttt{ocamlnat} prototype (as shown in
Figure~\ref{fig:ocamlnat_prototype}) with something that does not depend on any external programs
but does the executable code generation just-in-time within the process of the native toplevel.

\begin{figure}[htb]
  \centering
  \begin{tikzpicture}[node distance=8mm,text height=1.5ex,text depth=.25ex]
    \matrix[row sep={1.3cm,between origins}]
    {
      \node (start) {}; \\
      \node[phase] (compiler) {Native code compiler}; \\
      \node[phase] (jit) {Just-In-Time Emitter}; \\
      \node[phase] (jitld) {Just-In-Time Linker}; \\
      \node (end) {}; \\
    };
    \draw[->] (start) -- node[right] {\it OCaml phrase} (compiler);
    \draw[->] (compiler) -- node[right] {\it native code} (jit);
    \draw[->] (jit) -- node[right] {\it object code} (jitld);
    \draw[->] (jitld) -- node[right] {\it executable code} (end);
  \end{tikzpicture}
  \caption{\texttt{ocamlnat} overview}
  \label{fig:ocamlnat_overview}
\end{figure}

Figure~\ref{fig:ocamlnat_overview} shows our current implementation. We replaced
the Native code emitter and Toolchain Assembler phases from the \texttt{ocamlnat} prototype with
a Just-In-Time Emitter phase, and the Toolchain Linker and Runtime Linker phases with a Just-In-Time
Linker phase. The earlier phases, that are shared with the optimizing native code compiler
\texttt{ocamlopt} as described in Section~\ref{section:Overview_of_the_OCaml_compilers} and
\ref{section:The_native_toplevel}, remain unchanged.

The Just-In-Time Emitter phase is responsible for transforming the linearized native code that is
generated by the Native code compiler (as shown in Figure~\ref{fig:Native_code_generation}) into
\emph{object code} for the target platform. This \emph{object code} is very similar to the \emph{object
file} generated by the Toolchain Assembler; it contains a \texttt{text} section with the executable
code, a \texttt{data} section with the associated data items (\ie the floating-point constants, closures
and string literals used within the code, the frametable for the garbage collector, \ldots), a list
of relocations, and a list of global symbols.

The Just-In-Time Linker phase allocates executable memory for the \texttt{text} section and writable
memory for the \texttt{data} section, copies the section contents to their final memory locations,
takes care of the relocations, and registers the global symbols. This is roughly what the Toolchain
Linker and Runtime Linker in the \texttt{ocamlnat} prototype do.

The code for the two phases is found in the \texttt{toplevel/jitaux.ml} file, which provides the common, platform
independent functionality for Just-In-Time code generation, as well as \texttt{toplevel/amd64/jit.ml} for the x86-64
platform and \texttt{toplevel/i386/jit.ml} for the x86 platform, plus a few lines of additional C code in
\texttt{asmrun/natdynlink.c} and \texttt{asmrun/natjit.c}. At the time of this writing the changes for our
new native toplevel with support for x86 and x86-64 account for approximately 2300 lines of C and OCaml
code as shown in Table~\ref{tab:Additional_lines_of_code_for_ocamlnat}.

\begin{table}[htb]
  \centering
  \begin{tabular}{r|rr}
    & OCaml & C \\
    \hline
    Generic & $277$ & $185$ \\
    amd64 & $863$ & $\diagup$\, \\
    i386 & $991$ & $\diagup$\, \\
    \hline
    & $2131$ & $185$ \\
  \end{tabular}
  \caption{Additional lines of code for \texttt{ocamlnat}}
  \label{tab:Additional_lines_of_code_for_ocamlnat}
\end{table}

We tried to keep the code as easy to maintain as possible. To achieve this goal we
\begin{enumerate}[(a)]
\item reused as much of the existing functionality as possible of both the native code compiler and
  its runtime,
\item kept the amount of additional runtime support as low as possible (basically just an additional layer in
  the global symbol management and a new entry point for the Just-In-Time code execution), and
\item made the Just-In-Time Emitter (the \texttt{jit.ml} files in the \texttt{toplevel} subdirectories) look as
  similar as possible to the Native Code Emitter (the \texttt{emit.mlp} files in the \texttt{asmcomp}
  subdirectories).
\end{enumerate}
The last point is especially important as every change to the Native Code Emitters
(\ie \texttt{asmcomp/amd64/emit.mlp}) must be reflected by an equivalent change to the appropriate
Just-In-Time Emitter (\ie \texttt{toplevel/amd64/jit.ml}). Fortunately the Native Code Emitters usually
do not change very often during the OCaml development process.

\section{Performance} \label{section:Performance}

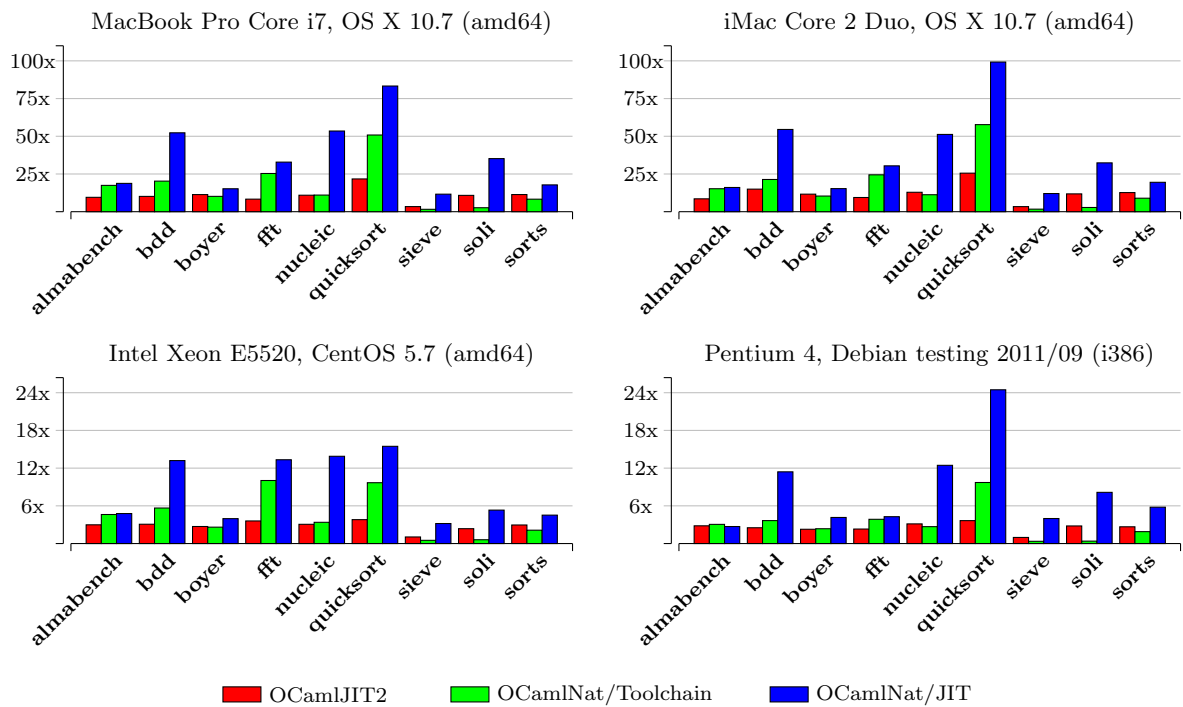
\begin{figure*}[bt]
  \centering
  \footnotesize
  \begin{tikzpicture}
    \begin{scope}[xshift=0cm,yshift=0cm] 
      \foreach \y in {1,...,4}  
      {
        \pgfmathtruncatemacro\ytext{\y * 25}
        \draw[gray!50, text=black] (-0.2 cm,\y * 0.5 cm) -- (6.6 cm,\y * 0.5 cm)
        node at (-0.5 cm,\y * 0.5 cm) {\ytext x};  
      };
      \draw (0cm,0cm) -- (6.6cm,0cm);  
      \draw (0cm,0cm) -- (0cm,-0.1cm);  
      \draw (6.6cm,0cm) -- (6.6cm,-0.1cm);  
      \draw (-0.1cm,0cm) -- (-0.1cm,2.2cm);  
      \draw (-0.1cm,0cm) -- (-0.2cm,0cm);  
      \draw (-0.1cm,2.2cm) -- (-0.2cm,2.2cm);  
      \node at (3.3cm, 2.5cm) {\small MacBook Pro Core i7, OS X 10.7 (amd64)};
      \foreach \name/\x/\a/\b/\c in
      {
        almabench/0/9.531/17.447/18.770,
        bdd/1/10.147/20.267/52.292,
        boyer/2/11.351/10.169/15.206,
        fft/3/8.323/25.330/32.869,
        nucleic/4/10.931/11.022/53.481,
        quicksort/5/21.714/50.833/83.298,
        sieve/6/3.382/1.590/11.625,
        soli/7/10.827/2.631/35.188,
        sorts/8/11.370/8.299/17.767
      }
      {
        \pgfmathsetmacro\xinc{0.2}
        \pgfmathsetmacro\xoff{\x * 0.7 + \xinc}
        \draw[fill=ocamljit2] (\xoff cm + 0 * \xinc cm, 0 cm) rectangle (\xoff cm + 1 * \xinc cm, \a * 0.02 cm);
        \draw[fill=ocamlnatext] (\xoff cm + 1 * \xinc cm, 0 cm) rectangle (\xoff cm + 2 * \xinc cm, \b * 0.02 cm);
        \draw[fill=ocamlnatjit] (\xoff cm + 2 * \xinc cm, 0 cm) rectangle (\xoff cm + 3 * \xinc cm, \c * 0.02 cm);
        \node[rotate=45,left] at (\xoff cm + 2.5 * \xinc cm, -.1cm) {\bf\name};
      };
    \end{scope}
    \begin{scope}[xshift=8cm,yshift=0cm] 
      \foreach \y in {1,...,4}  
      {
        \pgfmathtruncatemacro\ytext{\y * 25}
        \draw[gray!50, text=black] (-0.2 cm,\y * 0.5 cm) -- (6.6 cm,\y * 0.5 cm)
        node at (-0.5 cm,\y * 0.5 cm) {\ytext x};  
      };
      \draw (0cm,0cm) -- (6.6cm,0cm);  
      \draw (0cm,0cm) -- (0cm,-0.1cm);  
      \draw (6.6cm,0cm) -- (6.6cm,-0.1cm);  
      \draw (-0.1cm,0cm) -- (-0.1cm,2.2cm);  
      \draw (-0.1cm,0cm) -- (-0.2cm,0cm);  
      \draw (-0.1cm,2.2cm) -- (-0.2cm,2.2cm);  
      \node at (3.3cm, 2.5cm) {\small iMac Core 2 Duo, OS X 10.7 (amd64)};
      \foreach \name/\x/\a/\b/\c in
      {
        almabench/0/8.532/15.218/16.057,
        bdd/1/14.959/21.358/54.515,
        boyer/2/11.627/10.374/15.301,
        fft/3/9.416/24.440/30.396,
        nucleic/4/12.881/11.252/51.219,
        quicksort/5/25.551/57.709/99.199,
        sieve/6/3.360/1.690/12.042,
        soli/7/11.811/2.801/32.370,
        sorts/8/12.668/8.946/19.491
      }
      {
        \pgfmathsetmacro\xinc{0.2}
        \pgfmathsetmacro\xoff{\x * 0.7 + \xinc}
        \draw[fill=ocamljit2] (\xoff cm + 0 * \xinc cm, 0 cm) rectangle (\xoff cm + 1 * \xinc cm, \a * 0.02 cm);
        \draw[fill=ocamlnatext] (\xoff cm + 1 * \xinc cm, 0 cm) rectangle (\xoff cm + 2 * \xinc cm, \b * 0.02 cm);
        \draw[fill=ocamlnatjit] (\xoff cm + 2 * \xinc cm, 0 cm) rectangle (\xoff cm + 3 * \xinc cm, \c * 0.02 cm);
        \node[rotate=45,left] at (\xoff cm + 2.5 * \xinc cm, -.1cm) {\bf\name};
      };
    \end{scope}
    \begin{scope}[xshift=0cm,yshift=-4.4cm] 
      \foreach \y in {1,...,4}  
      {
        \pgfmathtruncatemacro\ytext{\y * 6}
        \draw[gray!50, text=black] (-0.2 cm,\y * 0.5 cm) -- (6.6 cm,\y * 0.5 cm)
        node at (-0.5 cm,\y * 0.5 cm) {\ytext x};  
      };
      \draw (0cm,0cm) -- (6.6cm,0cm);  
      \draw (0cm,0cm) -- (0cm,-0.1cm);  
      \draw (6.6cm,0cm) -- (6.6cm,-0.1cm);  
      \draw (-0.1cm,0cm) -- (-0.1cm,2.2cm);  
      \draw (-0.1cm,0cm) -- (-0.2cm,0cm);  
      \draw (-0.1cm,2.2cm) -- (-0.2cm,2.2cm);  
      \node at (3.3cm, 2.5cm) {\small Intel Xeon E5520, CentOS 5.7 (amd64)};
      \foreach \name/\x/\a/\b/\c in
      {
        almabench/0/2.994/4.641/4.793,
        bdd/1/3.087/5.677/13.254,
        boyer/2/2.725/2.620/3.980,
        fft/3/3.611/10.073/13.389,
        nucleic/4/3.076/3.396/13.944,
        quicksort/5/3.820/9.720/15.535,
        sieve/6/1.053/0.516/3.200,
        soli/7/2.375/0.609/5.344,
        sorts/8/2.967/2.136/4.542
      }
      {
        \pgfmathsetmacro\xinc{0.2}
        \pgfmathsetmacro\xoff{\x * 0.7 + \xinc}
        \draw[fill=ocamljit2] (\xoff cm + 0 * \xinc cm, 0 cm) rectangle (\xoff cm + 1 * \xinc cm, \a * 0.083 cm);
        \draw[fill=ocamlnatext] (\xoff cm + 1 * \xinc cm, 0 cm) rectangle (\xoff cm + 2 * \xinc cm, \b * 0.083 cm);
        \draw[fill=ocamlnatjit] (\xoff cm + 2 * \xinc cm, 0 cm) rectangle (\xoff cm + 3 * \xinc cm, \c * 0.083 cm);
        \node[rotate=45,left] at (\xoff cm + 2.5 * \xinc cm, -.1cm) {\bf\name};
      };
    \end{scope}
    \begin{scope}[xshift=8cm,yshift=-4.4cm] 
      \foreach \y in {1,...,4}  
      {
        \pgfmathtruncatemacro\ytext{\y * 6}
        \draw[gray!50, text=black] (-0.2 cm,\y * 0.5 cm) -- (6.6 cm,\y * 0.5 cm)
        node at (-0.5 cm,\y * 0.5 cm) {\ytext x};  
      };
      \draw (0cm,0cm) -- (6.6cm,0cm);  
      \draw (0cm,0cm) -- (0cm,-0.1cm);  
      \draw (6.6cm,0cm) -- (6.6cm,-0.1cm);  
      \draw (-0.1cm,0cm) -- (-0.1cm,2.2cm);  
      \draw (-0.1cm,0cm) -- (-0.2cm,0cm);  
      \draw (-0.1cm,2.2cm) -- (-0.2cm,2.2cm);  
      \node at (3.3cm, 2.5cm) {\small Pentium 4, Debian testing 2011/09 (i386)};
      \foreach \name/\x/\a/\b/\c in
      {
        almabench/0/2.835/3.065/2.721,
        bdd/1/2.525/3.667/11.459,
        boyer/2/2.284/2.367/4.170,
        fft/3/2.320/3.876/4.292,
        nucleic/4/3.142/2.700/12.499,
        quicksort/5/3.662/9.755/24.566,
        sieve/6/0.978/0.361/4.000,
        soli/7/2.813/0.388/8.182,
        sorts/8/2.678/1.903/5.812
      }
      {
        \pgfmathsetmacro\xinc{0.2}
        \pgfmathsetmacro\xoff{\x * 0.7 + \xinc}
        \draw[fill=ocamljit2] (\xoff cm + 0 * \xinc cm, 0 cm) rectangle (\xoff cm + 1 * \xinc cm, \a * 0.083 cm);
        \draw[fill=ocamlnatext] (\xoff cm + 1 * \xinc cm, 0 cm) rectangle (\xoff cm + 2 * \xinc cm, \b * 0.083 cm);
        \draw[fill=ocamlnatjit] (\xoff cm + 2 * \xinc cm, 0 cm) rectangle (\xoff cm + 3 * \xinc cm, \c * 0.083 cm);
        \node[rotate=45,left] at (\xoff cm + 2.5 * \xinc cm, -.1cm) {\bf\name};
      };
    \end{scope}
    \begin{scope}[xshift=2cm,yshift=-6.5cm]
      \draw[fill=ocamljit2] (0.0cm, 0.0cm) rectangle (0.5cm, 0.2cm) node[right] at (0.5cm, 0.1cm)
      {OCamlJIT2};
      \draw[fill=ocamlnatext] (3.0cm, 0.0cm) rectangle (3.5cm, 0.2cm) node[right] at (3.5cm, 0.1cm)
      {OCamlNat/Toolchain};
      \draw[fill=ocamlnatjit] (7.2cm, 0.0cm) rectangle (7.7cm, 0.2cm) node[right] at (7.7cm, 0.1cm)
      {OCamlNat/JIT};
    \end{scope}
  \end{tikzpicture}
  \caption{Speedup relative to the byte code toplevel \texttt{ocaml}}
  \label{fig:Speedup_relative_to_the_byte_code_toplevel_ocaml}
\end{figure*}

We compared the performance of our native toplevel to the performance of the byte code toplevel
\texttt{ocaml} running on top of the OCaml 3.12.1 byte code interpreter, the byte code toplevel
\texttt{ocaml} running on top of the OCamlJIT2 Just-In-Time byte code compiler
\cite{Meurer10jit,Meurer10ocamljit,Meurer11ocamljit2}, and Alain Frisch's earlier \texttt{ocamlnat}
proof-of-concept implementation (as described in Section~\ref{section:The_native_toplevel}).
 We measured the performance on four different systems:
\begin{itemize}
\item A MacBook Pro 13" (Early 2011) with an Intel Core i7 2.7GHz CPU (4 MiB L3 Cache, 256 KiB L2 Cache per Core,
  2 Cores) and 4 GiB RAM, running Mac OS X Lion 10.7.1. The C compiler is
  \texttt{llvm-gcc-4.2.1} (Based on Apple Inc. build 5658) (LLVM build 2336.1.00).
\item An iMac 20" (Early 2008) with an Intel Core 2 Duo ``Penryn'' 2.66GHz CPU (6 MiB L2 Cache, 2 Cores),
  and 4 GiB RAM, running Mac OS X Lion 10.7.1. The C compiler is
  \texttt{llvm-gcc-4.2.1} (Based on Apple Inc. build 5658) (LLVM build 2336.1.00).
\item A Fujitsu Siemens Primergy server with two Intel Xeon E5520 2.26GHz CPUs (8 MiB L2 Cache, 4 Cores),
  and 12 GiB RAM, running CentOS release 5.7 (Final) with Linux/x86\_64 2.6.18-274.3.1.el5.
  The C compiler is \texttt{gcc-4.1.2} (Red Hat 4.1.2-51).
\item A Fujitsu Siemens Primergy server with an Intel Pentium 4 ``Northwood'' 2.4 GHz CPU (512 KiB L2 Cache),
  and 768 MiB RAM, running Debian testing as of 2011/09 with Linux/i686 3.0.0-1-686-pae.
  The C compiler is \texttt{gcc-4.6.1} (Debian 4.6.1-4).
\end{itemize}

The OCaml distribution used for the tests is 3.12.1. The OCamlJIT2 version is the commit \texttt{8514ccb}
from the \texttt{ocamljit2} Git repository hosted on GitHub at \cite{Meurer11ocamljit2}. For
our native toplevel \texttt{ocamlnat} we used the commit \texttt{d30210d} from the \texttt{ocaml-experimental}
Git repository hosted on GitHub at \cite{Meurer11ocamlexperimental}.

The benchmark programs used to measure the performance are the following test programs
from the \texttt{testsuite/test} folder of the OCaml 3.12.1 distribution:
\begin{description}
\item[almabench] is a number-crunching benchmark designed for cross-language comparisons.
\item[bdd] is an implementation of binary decision diagrams, and therefore a good test for
  the symbolic computation performance.
\item[boyer] is a term manipulation benchmark.
\item[fft] is an implementation of the Fast Fourier Transformation \cite{BrighamM67}.
\item[nucleic] is another floating-point benchmark.
\item[quicksort] is an implementation of the well-known QuickSort algorithm \cite{Hoare61b,Hoare62}
  on arrays and serves as a good test for loops.
\item[sieve] is an implementation of the sieve of Eratosthenes, one of a number of prime number sieves,
  for finding all prime numbers up to a specified integer.
\item[soli] is a simple solitaire solver, well suited for testing the performance of non-trivial,
  short running programs.
\item[sorts] is a test bench for various sorting algorithms.
\end{description}

For our tests we measured the total execution time of the benchmark process itself and all spawned
child processes (only relevant for the earlier toolchain based \texttt{ocamlnat} prototype), given
as combined system and user CPU time. The times were collected by executing each benchmark two times
with every toplevel, and using the timings of the fastest run.

Figure~\ref{fig:Speedup_relative_to_the_byte_code_toplevel_ocaml} presents the results of the benchmarks
as speedup relative to the regular byte code toplevel \texttt{ocaml}, where OCamlNat/Toolchain is
Alain Frisch's toolchain based \texttt{ocamlnat} prototype and OCamlNat/JIT is our new Just-In-Time based
native toplevel. As you can see we managed to achieve speedups of up to hundred times faster than the
byte code toplevel in certain benchmarks. It is however worth noting that this is in part related to
the fact that \texttt{llvm-gcc} became the default C compiler with recent versions of OS X (and it's
related software development tools), which disables the very important manual register assignment
optimization in the byte code interpreter, because LLVM does not support manual register assignment.

\section{Related and further work} \label{section:Related_and_further_work}

We are looking forward to integrate our new native toplevel as interactive console
into the Mirage cloud operating system \cite{Mirage11,Madhavapeddy10,Madhavapeddy10hotcloud}, which
will be really useful for development (\ie when exploring the heap of Xen kernels interactively to
find resource leaks, \ldots).

Right now our \texttt{ocamlnat} implementation supports only x86 and x86-64 targets on POSIX systems.
We plan to add support for additional targets that are already supported by the optimizing native
code compiler, most notably ARM and PowerPC, and extend the support for other operating systems,
\ie Windows.

We are also working on integrating the linear scan register allocator \cite{PolettoS99,WimmerM05}
into the optimizing native code compiler as an alternative to the currently used graph coloring
register allocator \cite{Aho06}. By using the linear scan algorithm, which is commonly used in the
scope of Just-In-Time compilers, we expect to avoid the quadratic worst case running time of the
graph coloring algorithm\footnote{This work is also part of the first author's diploma thesis.}.

We are also in the process of evaluating the use of the LLVM compiler infrastructure
\cite{Lattner02,Lattner04,LLVM11} as a replacement
for the last phases of the native code compiler engine. Other projects such as Clang \cite{Clang11},
GHC \cite{Terei10} and MacRuby \cite{MacRuby11} already demonstrated the viability and usefulness of
using LLVM as compiler backend. Besides the other obvious benefits of using LLVM in OCaml, we would
also get the Just-In-Time compilation and execution engine for free.

\section{Conclusion} \label{section:Conclusion}

Our results demonstrate that an OCaml toplevel based on the native code compiler and runtime offers
significant performance improvements over the byte code toplevel (at least thrice as fast in all
benchmarks, and up to hundred times faster in one benchmark), at acceptable maintenance costs.

As demonstrated in Section~\ref{section:Performance} we were also able to beat the performance of
our earlier byte code based OCamlJIT2 prototype in almost every case (except for the \texttt{almabench}
benchmark on x86, which is due to the fact that OCamlJIT2 uses SSE2 registers and instructions while
our native toplevel uses the x87 FPU stack and instructions \cite{Meurer10jit}).

\section*{Acknowledgements} \label{section:Acknowledgements}

We would like to thank Alain Frisch and Fabrice Le Fessant for sharing their earlier work on
in-process object code generation with us, which inspired our current work to some degree.
We would also like to thank Simon Meurer for his careful proof-reading.

\bibliographystyle{abbrv}
\bibliography{citations}

\end{document}